\documentclass [twocolumn,showpacs,preprintnumbers,prl,tightenlines,
  superscriptaddress,footinbib]{revtex4} \usepackage{amssymb}
\usepackage{graphicx} \usepackage{dcolumn} \usepackage{bm}
\usepackage{amsmath} \usepackage{amsfonts}%
\providecommand{\U}[1]{\protect\rule{.1in}{.1in}}
\providecommand{\U}[1]{\protect\rule{.1in}{.1in}} \usepackage{epsfig}
\newcommand{\be}{\begin{equation}} \newcommand{\ee}{\end{equation}}
\newcommand{\bea}{\begin{eqnarray}} \newcommand{\eea}{\end{eqnarray}}

\begin{document}

\title{Griffiths phases on complex networks}

\author{Miguel A. Mu\~noz} \affiliation{Institute \textit{Carlos I}
  for Theoretical and Comp. Physics. Universidad de Granada, 18071
  Granada, Spain.}  \author{R\'obert Juh\'asz}
\affiliation{Research Institute for Solid
State Physics and Optics, H-1525 Budapest, P.O.Box 49, Hungary}
\author{Claudio Castellano} 
\affiliation{CNR-ISC, Unit\`a SMC and Dip. di  Fisica, Univ. di Roma
  ``La Sapienza'', P. A. Moro 2, 00185 Roma, Italy} \author{G\'eza
  \'Odor}
\affiliation{Research Institute for Technical Physics and Materials
  Science, H-1525 Budapest, P.O.Box 49, Hungary} 


\begin{abstract} 
Quenched disorder is known to play a relevant role in dynamical processes and
phase transitions. Its effects on the dynamics of complex networks have hardly
been studied. Aimed at filling this gap, we analyze the Contact Process,
i.e. the simplest propagation model, with quenched disorder on complex
networks. We find Griffiths phases and other rare region effects, leading
rather generically to anomalously slow (algebraic, logarithmic, ...)
relaxation, on Erd\H os-R\'enyi networks. Similar effects are predicted to
exist for other topologies with a finite percolation threshold.  More
surprisingly, we find that Griffiths phases can also emerge in the absence of
quenched disorder, as a consequence of topological heterogeneity in networks
with finite topological dimension. These results have a broad spectrum of
implications for propagation phenomena and other dynamical processes on
networks.
\end{abstract}
\maketitle

Networks have become a paradigm in the study of complex systems
\cite{Laszlo}.  After initial efforts devoted to uncover their
non-trivial topological features, the focus shifted to dynamical
processes occurring on them \cite{Porto}. Models of epidemics
\cite{AS}, such as the susceptible -infected -recovered (SIR), the
susceptible -infected -susceptible (SIS), or the contact process (CP),
have played a prominent role \cite{Porto}. The concept of
``epidemics'' covers a broad variety of processes including real
epidemics, computer viruses, rumor spreading, or signal propagation in
neural nets. A remarkable finding is the absence of a finite infection
threshold for the SIS process in heterogeneous scale-free networks;
i.e.  in contrast with regular lattices, where a critical point
separates an active from an absorbing phase \cite{AS}, these networks
host endemic states even for arbitrarily small infection rates
\cite{Porto}; i.e.  {\it topological disorder} crucially affects
dynamical processes on networks.

{\it Quenched disorder} is well known to induce novel behavior (such as,
universality changes) in phase transitions both in equilibrium and away from
it.  Under some circumstances, it may also generate phases unheard-of in pure
systems.  Consider, for illustration, the quenched contact process (QCP)
\cite{QCP}, characterized by a site-dependent quenched infection rate, on a
lattice. Even if the system is globally in its absorbing phase, disorder
fluctuations can generate {\it rare active regions} with over-average
infection rate. In these, activity lingers for extremely long periods;
however, being finite, they ineluctably end up falling into the absorbing
state.  The convolution of exponentially rare regions with exponentially large
surviving times gives rise (see below) to a region in the absorbing phase,
characterized by a generic algebraic decay of activity, i.e. a {\it Griffiths
  phase} (GP) \cite{GP,QCP,Vojta}.

This is just an example of a generic phenomenon, thoroughly studied in the
disordered systems literature in classical, quantum, and non-equilibrium
systems \cite{GP}, occurring whenever exponentially distributed cluster sizes,
$P(s)\sim \exp(- c s)$, have exponentially long activity times, $\tau(s) \sim
\exp(b s)$; this leads to a GP with algebraic decay $\rho~\sim \tau^{-c/b}$
with a continuously varying exponent (see \cite{GP} for analytical approaches,
further details and applications, and \cite{Vojta} for a recent review).
Observe that, on the opposite to usual critical points, in GPs one observes
scale-invariance generically, i.e.  without the need of parameter fine-tuning.
Similarly, other size distributions (as power-laws) do lead to different
functional forms of slow relaxation.  Let us stress that many disorder effects
appear rather similarly in equilibrium and away from it; for instance, the
(strong disorder) critical point of the (non-equilibrium) QCP is in the same
universality class as the (equilibrium) random transverse-field Ising model
\cite{Hoo}; i.e. rare-region effects transcend the frontier between
equilibrium and non-equilibrium.

From this broad perspective, it is surprising that very little attention has
been paid so far to the effect of quenched disorder on dynamical processes on
complex networks. Heterogeneity in the intrinsic properties of nodes is a very
natural (not to say, unavoidable) feature of real networks: node-dependent
rates appear in all the examples above, owing to the specificity of the
individual immune response, presence of antivirus software, and so forth. Our
aim here is twofold: (i) we tackle the study of quenched node disorder in
dynamical processes on complex networks, and look for rare-region effects in
the simplest possible epidemic model, i.e. the QCP on Erd\H os-R\'enyi (ER)
random networks; (ii) we explore whether network topological disorder on its
own can induce rare-region effects. We report on the existence of Griffiths
effects, including various non-trivial regimes with generic slow decay of
activity, for different networks with quenched node disorder and/or
topological disorder. Our main conclusion is that disorder, either quenched or
topological, may induce under the conditions studied here, slow relaxation on
network dynamics; this is expected to go beyond the considered examples, and
to apply to different models, dynamics and topologies.

In the {\it pure} Contact Process, infected/active nodes heal at rate $\mu$
and infect a randomly chosen nearest-neighbor (provided it was un-infected) at
rate $\lambda$.  A simple (one-site) mean-field treatment of the average
activity density, $\rho$, predicts an absorbing phase transition at
$\lambda_c^{1}=\mu$, i.e.  where the infection and healing rate compensate
each other \cite{Castellano06}.
This prediction for the threshold value is not correct for {\it finitely
  connected} networks since in that case activity appears in localized
regions, reducing the effective infection rate (i.e. the prob. of having an
already infected neighbor is larger than in the perfect mean-field mixing),
i.e. $\lambda_c> \lambda_c^{1} $.  A lengthy but standard \cite{AS} two-site
or ``pair'' approximation, leads to $\lambda_c^{2}= \mu \langle k \rangle
/(\langle k \rangle-1) $, where $\langle k \rangle$ is the average network
degree \cite{Long}.  Observe that the correction factor with respect to
$\lambda_c^{1}$ converges to $1$ when $\langle k \rangle \rightarrow \infty$
(i.e. where mean-field holds) and diverges at the percolation threshold
$\langle k \rangle=1$ below which the network is fragmented
\cite{Bollobas,Porto}, it cannot sustain activity and the phase transition
disappears.

We consider a {\it QCP} \cite{QCP} with $\mu=1$ and a quenched
disordered infection rate: a fraction $1-q$ of the nodes (type-I) take
value $\lambda$ and the remaining fraction $q$ (type-II nodes) take a
reduced value $\lambda r$, with $0 \le r < 1$.  Obviously, for $q=0$
and $q=1$ the model is pure, $\lambda_c(q=1)= \lambda_{c}(q=0)/r$,
while for $0<q<1$ one expects $\lambda_{c}$ to interpolate between
these limits. For simplicity, we fix $r=0$ from now on. The average
activity density is $\rho = (1- q) \rho_1 + q \rho_2$ (subscripts
standing for node type) and the one-site mean-field equations are
$\dot{\rho}_i(t) = - \rho_i +\lambda (1- \rho_1 - \rho_2) ((1-q)
\rho_1)$
for $i=1, 2$, whose stationary solution has a critical point at
$\lambda_c^{1}(q) =1/(1-q)$.  As above, the one-site result needs to
be corrected by a factor $\langle k \rangle/(\langle k \rangle-1)$ to
account for activity clustering, leading to:
\begin{equation}
\lambda_{c}^{2}(q) = \frac{\langle k \rangle}{\langle k \rangle-1} ~
\frac{1}{1-q}.
\label{Crit}
\end{equation}
Relevant for what follows is that type-I nodes experience a percolation
transition where the type~I-to-type~I average degree is $1$
\cite{Bollobas,Porto}, i.e. at $q_{perc}= 1-\langle k \rangle^{-1}$. For $q
>q_{perc}$ activity cannot be sustained: type-I clusters are finite and
type-II ones do not propagate activity.

\begin{center}
\begin{figure}
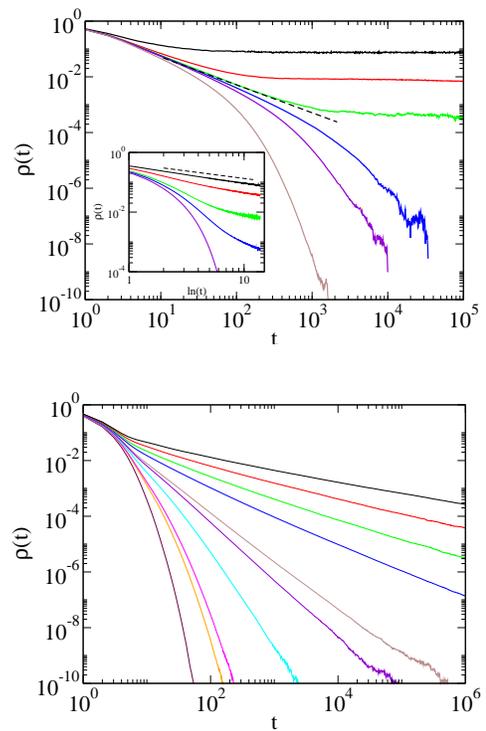

\includegraphics[height=4.5cm,angle=0]{Fig1a.eps}
\\
\vspace{0.57cm}
\includegraphics[height=4.5cm,angle=0]{Fig1b.eps}
\caption{Average activity density $\rho(t)$ vs time $t$ for ER
  networks with $\langle k \rangle = 3$, $r=0$, and $N=10^5$.
  $\lambda$s are ordered from top to bottom in all panels.  {\bf (a)
    Upper panel}: $q=0.6$, and
$\lambda=5, 3.8, 3.6, 3.55, 3.5, 3.3$; the dashed line is proportional to
  $t^{-1}$.  {\bf (a) Inset}: $\rho$ vs $\ln(t)$ for $q=2/3$;
$\lambda=10, 7, 5, 4.5, 4$; the dashed line is proportional to
  $\ln(t)^{-1/2}$.  {\bf (b) Lower panel}: $q=0.9$, and
  $\lambda=50,30,20,15,10,9,7,5,4.5,2.7$. Straight lines lie in the
Griffiths phase.
}
\label{numerics}
\end{figure}
\end{center}
\begin{center}
\begin{figure}
\includegraphics[height=5.5cm,angle=0]{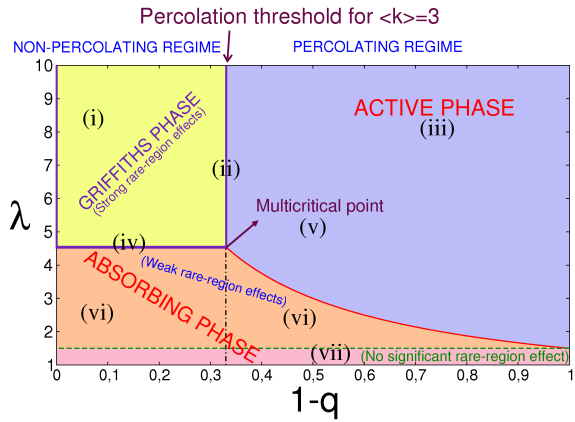}
\caption {Phase diagram for $r=0$. See main text for details.}
\label{Schema}
\end{figure}
\end{center}

\vspace{-1.00cm} We now present the results of a numerical investigation of
the QCP on ER networks with $\langle k \rangle = 3$ (implying $q_{perc}=2/3$),
and sizes up to $N=10^7$. Simulations are performed in a standard way: all
sites are declared active initially and the dynamics proceeds as follows
\cite{AS}. A site, $i$, is randomly selected and it either heals (with prob.
$1/(1+\lambda_i)$) or infects a randomly selected neighbor provided it was
empty (with prob. $\lambda_i/(1+\lambda_i)$, where $\lambda_i= \lambda$ or $0$
depending on site type). We monitor the activity decay averaged over many
runs.  Numerical results are synthesized in Fig.~\ref{numerics} and the
inferred phase-diagram is summarized in Fig.~\ref{Schema}. For $q<2/3$ a
critical line $\lambda_c(q)$, which is very well fitted by Eq.~(\ref{Crit})
(implying that the pair approximation captures the most relevant
correlations), separates an active phase from an absorbing one
(Fig.~\ref{numerics}a).  Instead for the fragmented case, $q>2/3$ there is no
active phase, as predicted above (Fig.~\ref{numerics}b).  The absorbing phase
can be divided into various {\it sub-phases}.  For $q > 2/3$ and $\lambda(q) >
\lambda_c(q=q_{perc}) \approx 4.5$ we observe (Fig.~\ref{numerics}b) a {\bf
  power-law decay with continuously varying exponents}: $\rho(t) \sim
t^{-\theta(q,\lambda)}$, with $\theta \rightarrow 0$ $q \rightarrow q=2/3$
i.e. a GP. Right at $q=2/3$ we find a {\bf logarithmic} decay $\rho(t) \sim
\ln(t)^{-1/2}$ for $\lambda > \lambda_c(q_{perc}) \approx 4.5$
(Fig.~\ref{numerics}a (inset)).  For $q > 2/3$ and $\lambda_c(q=0) < \lambda <
\lambda_c(q_{perc})$, as well as for $q<2/3$ and $\lambda_c(q=0)<
\lambda<\lambda_c(q)$ (see Fig.~\ref{Schema}), we can fit a {\bf stretched
  exponential} (not shown).
Finally, below the threshold of the pure system $\lambda_c(q=0) \approx 1.5$
the decay is purely exponential.
\vspace{-0.75cm} Let us now rationalize these observations by using, as
customarily done in disordered systems (see \cite{QCP,GP,Lee,Vojta}), {\it
  optimal fluctuation arguments}.  The following regimes can be predicted:
{\bf (i)} {\it Griffiths phase}: $\lambda > \lambda_c(q_{perc})$ and $q >
q_{perc}$.  The cluster size distribution for ER networks in the fragmented
case is \cite{perc-er}:
\begin{equation}
P(s) \sim \frac{1}{\sqrt{2 \pi} p} s^{-3/2} e^{-s(p-1-\ln(p))}
\label{size}
\end{equation}
where $s$ is the cluster size and $p$ is the average number of links per node,
which in our case is $p=\langle k \rangle_{q=0} (1-q)$ for type-I
nodes. Within any given cluster of type-I nodes, let us define $p_{loc}$ as
the local average number of links per node; and from it, a local value of $q$,
$q_{loc}= 1- p_{loc}/\langle k \rangle_{q=0}$.  Obviously, for connected
type-I clusters are above the percolation threshold, i.e. $q_{loc}<q_{perc}$,
and hence, if $\lambda > \lambda_c(q_{perc})$, they are active rare regions
where activity survives until a coherent fluctuation extinguishes it.  This
occurs after a characteristic time $\tau(s)$ which grows exponentially with
cluster size, i.e.  $\tau\simeq t_0 \exp(A(\lambda) s)$, where $t_0$ and
$A(\lambda)$ do not depend on $s$.  Hence, the overall activity decays as:
\begin{equation}
 \rho(t) \sim \int ds ~s ~P(s) \exp{[-t/(t_0 e^{A(\lambda) s})]}.
\label{convolution}
\end{equation} 
Plugging Eq.~(\ref{size}) into Eq.~(\ref{convolution}) and using a saddle
point approximation, one obtains $\rho(t) \sim t^{-\theta(p,\lambda)}$, with
$\theta(p,\lambda) = - (p-1-\ln(p))/ A(\lambda) $; i.e. there is a {\bf
  generic power-law decay with continuously varying exponents}, i.e. a
GP. Note that $\theta \rightarrow 0$ when $p \rightarrow 1$, suggesting a
crossover to logarithmic decay.  {\bf (ii)} Indeed, right at the percolation
threshold, $q=q_{perc}$, the cluster size distribution, Eq.~(\ref{size}),
becomes a power-law leading, when plugged in Eq.~(\ref{convolution}), to a
{\bf logarithmic decay}, $\rho(t) \sim [\ln(t/t_0)]^{-1/2}$, for any $\lambda$
larger than $\lambda_c(q_{perc})$ (for which rare regions are active).  {\bf
  (iii)} For $q<q_{perc}$ there is a giant component of type-I nodes and,
hence, standard mean-field like contact-process behavior is expected; i.e.
{\bf power-law decay}, $\rho(t) \sim t^{-1}$ at criticality.
Similar arguments lead to:
  { \bf (iv)} In the absorbing region with $\lambda_c(q=0) < \lambda <
  \lambda_c(q_{perc})$, rare regions with $q_{loc}<q_{perc}$ exist but, in
  contrast to previous cases, they are subcritical (activity decays
  exponentially fast in each of them); a {\bf stretched exponential} fits the
  decay. {\it Active} rare clusters can also appear; but they require
  $q_{loc}$ to be smaller that the critical value of $q$ for the given
  $\lambda$,
a stringent condition, leading (in all the region {\bf (iv)} in
Fig.(\ref{Schema}))
to {\it weak rare-region effects} (as opposed to the
previous {\it strong effects}) 
\cite{Long}.  Finally, region {\bf (v)} with $\lambda < \lambda_c(q=0) $ is
free from rare-region effects: all clusters are subcritical with {\bf
  exponential decay}.

Summing up, optimal fluctuation arguments explain all numerical findings.
{\it Rare regions play a key role giving rise to generic slow decay of
  activity}. Similar results hold for any value of $r$, i.e. when activity
flows out of any site. Hence, strong rare-region effects are expected to occur
in generic networks with a finite percolation threshold (e.g. structured
scale-free networks \cite{Structured} or networks with communities), while if
the percolation threshold vanishes (e.g. Barab\'asi-Albert scale-free nets
\cite{Laszlo}) only weak rare-region effects are predicted.

\begin{center}
\begin{figure}
\includegraphics[height=5.5cm,angle=0]{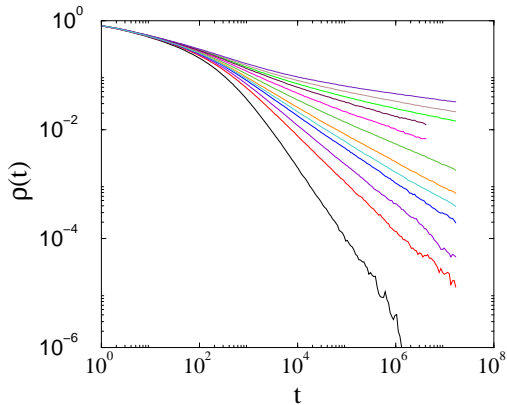}
\caption{Density decay in generalized small-world networks with $\alpha=2$ and
  $\beta=0.2$ for different values of $\lambda$ (from top to bottom: $2.81,
  2.795, 2.782, 2.77, 2.75, 2.73, 2.71, 2.70$, $2.69, 2.67, 2.65, 2.6$).
  Straight lines lie in the Griffiths phase.
}
\label{BB}
\end{figure}
\end{center}

In what follows, we investigate whether GPs induced merely by topological
disorder can exist. Let us first recall that Eq.(\ref{Crit}) provides an
excellent approximation to the ($\langle k \rangle$-dependent) critical
point. As a consequence, networks with heterogeneous degree may, in principle,
exhibit {\it topological rare regions}: clusters with local degree, $k$, above
average would have a smaller percolation threshold and could be locally active
even if the whole network is absorbing.  For argument's sake, let us consider
the CP on a network with bimodal degree distribution, $P(k) = p \delta(k-k_1)
+ (1-p) \delta(k-k_2) $ with $k_1 \gg k_2$; a priori, one could expect rare
active regions (with over-density of $k_1$-nodes) to exist.  However,
numerically we find just conventional, non-disordered, exponential decay.  Why
is it so?  In $d$-dimensional lattices disorder is known to be irrelevant for
large $d$; the number of nearest neighbors is so high that the central limit
theorem precludes rare regions (i.e. deviations from normality) from existing.
An extension of the concept of Euclidean dimension to arbitrary graphs is the
{\it topological dimension}, $D$, which measures how the number of nodes in a
neighborhood grows as a function of the topological distance from an arbitrary
origin: $ N(l) \sim l^{D}$~\cite{Bollobas}.  In parallel with the result for
lattices, we conjecture that GPs and similar rare-region effects do not
typically exist for the CP in networks with an infinite topological dimension
(such as ER graphs {\it above} their percolation threshold or the discussed
bimodal graphs), justifying our numerical findings. Instead, for ER {\it
  below} the percolation threshold, which are characterized by a vanishing
effective topological dimension, $D=0$ (i.e. the number of nodes in any
neighborhood/cluster converges to a constant for large values of $l$) GPs can
exist. This suggests to explore CP on networks with finite $D$. As an example,
we take the generalized random {\it small-world} networks defined in \cite{bb}
as follows. Starting with a ring of $L$ nodes, one defines the distance
between nodes $i$ and $j$ as $l=\min (|i-j|,L-|i-j|)$; all nearest neighbors
are connected and any pair with $l>1$ is connected with probability
$P(l)=1-\exp (-\beta l^{-\alpha})$. We have identified three different cases:
{\bf i}) For $\alpha < 2$ the network diameter grows poly-logarithmically with
$N$, hence, formally $D =\infty$; indeed, as conjectured, no Griffiths phases
are observed in our numerics (not shown). The critical behavior seems to
correspond to that of the contact process with L\'evy-flight jumps
\cite{AS}. {\bf ii}) For $\alpha > 2$, $\langle l \rangle$ is finite, and
correspondingly $D=1$; long-edges do not alter the system dimensionality but
they {\it do} introduce quenched disorder and a GP. We conjecture the critical
behavior to coincide with that of the one-dimensional infinite randomness
fixed point of the quenched contact process \cite{Vojta,Hoo}, as indeed
verified numerically (not shown); topological and quenched disorder can lead
to the same universal behavior.  {\bf iii}) In the marginal case $\alpha=2$,
$D$ has been conjectured to be $\beta$-dependent, $D=D(\beta)$
\cite{bb,Juhasz}.  We have checked that $D$ is indeed an increasing function
of $\beta$ ($D(0)=1$, $D(0.2)\approx 1.21$, $D(0.5)\approx 2.3$).  We expect
the critical point $\lambda_c(\beta)$ to decrease when $\beta$ grows
(i.e. when $\langle k \rangle$ increases; see Eq.(\ref{Crit})). Hence, for a
given $\beta$, the model must be in the active phase if
$\lambda>\lambda_c(0)=3.297848(22)$ (critical point in the one-dimensional
lattice \cite{AS}) and must be inactive if
$\lambda<\lim_{\beta\to\infty}\lambda_c(\beta)=1$ (critical point for the
fully connected graph).  Therefore, $\lambda_c(\infty) \le\lambda_c(\beta) \le
\lambda_c(0) $ and the possible GP is bounded by this interval. Numerical
simulations for $\beta=0.2$ confirm (Fig.~\ref{BB}) the existence of a GP with
generic power-law decay for $\lambda$ in $[2.65,2.81]$.  The width of the GP
decreases with increasing $\beta$ (i.e. the larger $D$ the smaller the rare
region effects); actually, preliminary results, suggests the existence of a
finite upper critical dimension above which the GP disappears \cite{Long}.  We
have found similar topological GPs for other small-world nets, and we predict
them
to emerge in many other networks, such as spatially embedded networks or
fractal \cite{Makse} scale-free ones, with finite topological dimension.
\vspace{-0.75cm} In summary, taking the contact process as an example we have
shown that quenched disorder can induce GPs and other rare-region effects
leading to generic slow relaxation for dynamical processes on ER
networks. Similar effects are argued to appear on other topologies with a
finite percolation threshold. Furthermore, heterogeneity in the topology may
suffice to generate GPs on its own. This can occur only if the network
topological dimension is not infinite allowing for rare regions to exist. Our
results are expected to apply to dynamical processes other than the CP (even
if each case should be carefully examined).  An inspiring application,
illustrating Griffiths effects at play, is provided by a recent work reporting
on algebraic ``forgetting'' times (reproducing experimental findings) in a
simple model of memory \cite{Johnson}; this generic slow decay crucially
depends on the neural network topological disorder. Social networks with
heterogeneous communities have also reported to exhibit generic slow decay
and, what we interpret as, severe rare-region effects \cite{Castello}.

This work was supported by HPC-EUROPA2 pr.228398,
HUNGRID and Hungarian OTKA (T77629,K75324), J. de Andaluc{\'\i}a
P09-FQM4682 and MICINN--FEDER project FIS2009--08451.

\end{document}